\documentclass[dvips]{article}
\usepackage{icrctc07}

\title{Long term monitoring of bright TeV Blazars with the MAGIC telescope}
\shorttitle{Blazar Monitoring with MAGIC}

\authors{F. Goebel$^{1}$, M. Backes$^2$, T. Bretz$^3$,
M. Hayashida$^{1}$, C. Hsu$^{1}$, K. Mannheim$^3$, A. Moralejo$^4$, W. Rhode$^2$,
K. Satalecka$^{4}$, M. Shayduk$^{1,5}$, M. Teshima$^{1}$,
R. M. Wagner$^{1}$\\ for the MAGIC collaboration}
\shortauthors{Florian Goebel et al.}
\afiliations{$^1$Max-Planck-Institut f\"ur Physik, Munich, Germany \\ 
  $^2$Universit\"at Dortmund, Germany \\ 
  $^3$Institut f\"ur Theoretische Physik und Astrophysik, Universit\"at W\"urzburg, Germany \\
  $^4$Institut de Fisica d'Altes Energies (IFAE), Barcelona, Spain \\
  $^5$DESY, Zeuthen, Germany \\
  $^6$Humboldt Universit\"at zu Berlin, Germany \\
}
\email{fgoebel@mppmu.mpg.de}

\abstract{The MAGIC telescope has performed long term monitoring observations of
the bright TeV Blazars Mrk421, Mrk501 and 1ES1959+650. Up to 40
observations, 30 to 60  minutes each have been performed for each source
evenly distributed over the observable period of the year. The
sensitivity of MAGIC is sufficient to establish a flux level of 25\% of
the Crab flux for each measurement. These observations are well suited
to trigger multiwavelength ToO observations and the overall collected
data allow an unbiased study of the flaring statistics of the
observed AGNs.}

\begin{document}
\maketitle
\section{Introduction}

The extra-galactic GeV/TeV $\gamma$-ray sources Mrk421, Mrk501 and 1ES1959+650
are blazars, i.e. Active Galactic Nuclei (AGNs) containing jets
(plasma outflows moving at relativistic velocities) pointing towards the
Earth. They belong to the subclass of High frequency-peaked BL Lac (HBL)
objects. These blazars exhibit no emission lines but a continuous Spectral
Energy Distribution (SED) with a peak in the UV to soft X-ray band and
a second peak in the GeV-TeV range.
These objects show prominent time variabilities on various time scales
at all frequencies. In the Very High Energy (VHE: 100~GeV - 100~TeV) range
variations larger than one order of magnitude and flux doubling times
of less than 10 minutes have been observed. 

Synchrotron-Self Compton (SSC) models,
which attribute the low energy
peak to Synchrotron radiation of relativistic electrons and the high
energy peak to inverse Compton radiation of the same electron
population, have been successfully used to describe most of the
existing multiwavelength data. However, the existing data are not
sufficient to distinguish between different models describing the
emission processes and the formation and structure of the jets. In
particular models in which hadronic acceleration plays a decisive role
are an attractive alternative. These models predict emission of high
energy neutrinos and may be more suitable than simple SSC models to
explain the observed orphan TeV flares of 1ES1959+650.

The sources Mrk421 (z=0.030), Mrk501 (z=0.034) and 1ES1959+650
(z=0.047) are bright and close blazars and therefore well suited
to study the intrinsic properties of these objects.

\section{AGN monitoring}

Monitoring the variable flux states of AGNs in VHE $\gamma$-rays using
Imaging Atmospheric Cherenkov Telescopes (IACTs) is in many ways a
valuable tool to study the jet physics of AGNs.
The measurement of the long term flux variability of blazars
is interesting in its own right and can provide input to
constrain theoretical models. The determination of flaring
state probabilities is also essential to estimate the statistical
significance of possible correlations between flaring states and other
observables such as neutrino events \cite{neutrino, neutrino2}. The selected AGNs are in the FoV
and some of the prime targets of the neutrino observatory
Amanda/IceCube which is observing the northern hemisphere with a yearly
improving sensitivity. 

In order to obtain an unbiased distribution of the flux level states
it is important to schedule the monitoring observations used for these
studies independently of any a priori knowledge of the flux
state. In particular observations triggered by observed high flux states in
the X-ray or $\gamma$-ray band should not be used for such studies.

Very importantly, AGN monitoring also allows one to trigger Target of
Opportunity (ToO) observations which require a high flux level in the
VHE range. Observations during high flux states are particularly
interesting since on the one hand these correspond to the most violent
states of the AGNs and on the other hand the high flux levels allow
very precise and high statistics observations.  
The ToO observations may be performed by the same IACT issuing the ToO
trigger but may also include other IACTs in order to increase the
time coverage of the observations. The ToO may also include
multiwavelength observations e.g. together with X-ray satellites.

The usual procedure to trigger AGN flare ToOs relies on X-ray
monitoring. This procedure has a few disadvantages, besides
obvious technical advantages. Although a general correlation between
X-ray and $\gamma$-ray flares cannot be denied, a strategy which only
relies on X-ray triggers is biased and will never detect the very interesting
orphan flares, which are characterized by high $\gamma$-ray states without
a simultaneous high state in X-rays.

Finally, the combined data obtained during unbiased monitoring observations can
be used to perform detailed AGN studies which require high statistics
at various flux levels. In particular VHE observations of AGNs during
low flux states are still rare.

\section{AGN monitoring strategies}

VHE $\gamma$-ray astronomy is currently a very dynamic field with many new
detections of exciting galactic and extra-galactic objects every
year. The observation time of IACTs is thus very precious and can
only to a small extent be devoted to AGN monitoring programs. 
Previous generation IACT which are still operational have therefore
been used to continuously monitor known AGNs~\cite{Whipple_monitoring} and
dedicated, small inexpensive IACT are under discussion.

Here we present a monitoring program using the MAGIC telescope, a high
sensitivity latest generation IACT. Short observations are scheduled
evenly distributed over the observable period of the year. Each of
these sampling observations should be long enough to detect a given
minimum flux level, taking into account the sensitivity of the
telescope. Typically, 20 to 60 min observations are sufficient to
detect moderate flaring states of nearby TeV blazars. These
observation times are short enough to keep the impact on the overall
observation schedule low. In the case of MAGIC, observations can be
scheduled during partial moon or modest twilight. This further
decreases the impact on high priority, deep observations, while the
sensitivity under these observational conditions is only slight
reduced~\cite{MAGIC_moon} for the purposes of the monitoring program.

Continuous observation programs with moderate sensitivity IACTs are
only sensitive to variations of the flux level averaged over longer
observation times. On the other hand short sampling observations with
high sensitivity IACTs are sensitive to considerably shorter flares
but the duty cycle is much lower.

\section{AGN monitoring program using the MAGIC telescope}

\begin{figure}
  \centering
  \includegraphics[width=7.5cm]{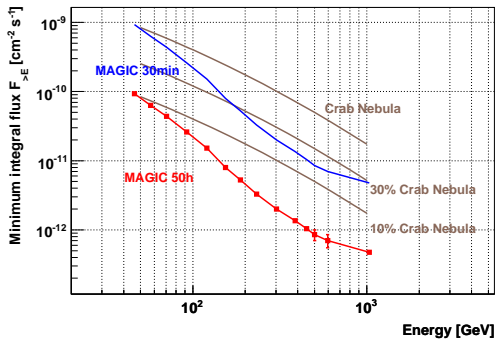}
  \caption{Sensitivity of MAGIC for 50 h and 30 min observations}\label{fig:MAGIC_sensi}
\end{figure}

MAGIC~\cite{MAGIC} is currently the largest single dish Imaging
Atmospheric Cherenkov telescope (IACT) for high energy $\gamma$-ray
astronomy with the lowest energy threshold among existing IACTs. It is
installed at the Roque de los Muchachos on the Canary Island La Palma
at 2200 m altitude and has been in scientific operation since summer
2004. The 17 m diameter parabolic shaped mirror preserves the
time structure of the Cherenkov light signals. The camera is equipped
with 576 photo-multiplier tubes (PMTs). The analog signals are
transfered via optical fibers to the trigger and FADC electronics. 
The energy threshold of MAGIC is $\sim$60~GeV. A source
emitting $\gamma$-rays at a flux level of 2.5\% of the Crab Nebula can be
detected with 5 sigma significance within 50 h observation time.
The sensitivity is sufficient to establish a flux level of 25\% of
the Crab flux above 300~GeV for a 20~min observation.
A quick online analysis estimates the flux level of each source during
data taking. The sensitivity of the online analysis allows to detect
a flux of 30\% of the Crab flux within 30 min (see
Figure~\ref{fig:MAGIC_sensi}).

The bright TeV blazars Mrk421, Mrk501 and 1ES1959+650 haven been
selected for a long term monitoring program with the MAGIC telescope
using the above mentioned sampling strategy. Up to 40 short
observations, evenly distributed over the respective observable time
during the MAGIC Cycle II observation period, have been scheduled of
each of these sources. For the brighter sources Mrk421 and Mrk501 30
min have been scheduled for each observation while for the less bright
blazar 1ES1959+650 60 min have been scheduled. The observation times
are enough to establish a flare using the online analysis and
can be used to trigger ToO observations.

\section{First results}

\begin{figure*}
  \centering
  \includegraphics[width=14cm,height=5.5cm]{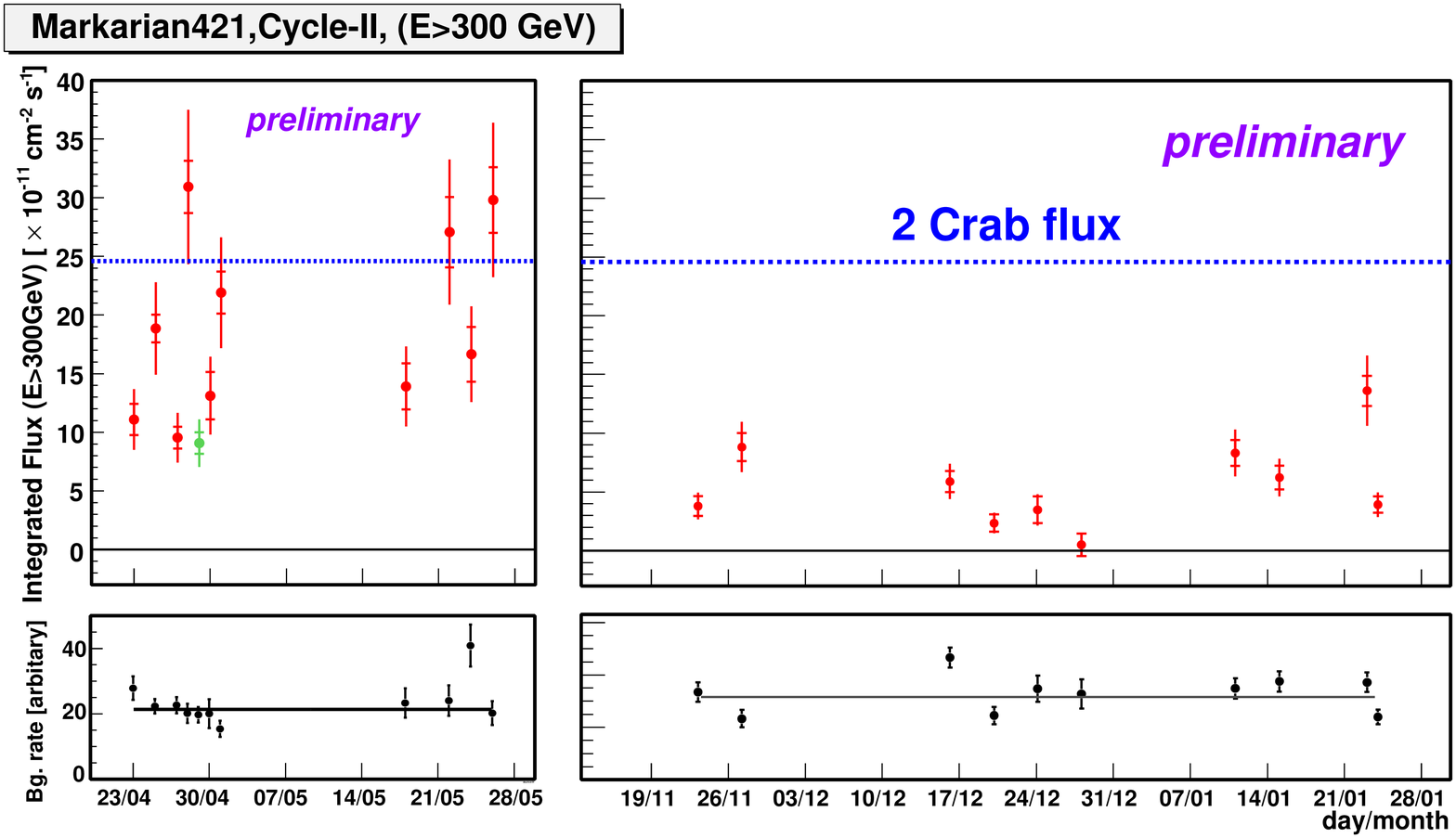}
  \caption{Light curves of Mrk421 observed between April 2006 and
    January 2007}\label{fig:Mrk421_lc}
\end{figure*}

\begin{figure*}
  \centering
  \includegraphics[width=14cm,height=6cm,angle=0]{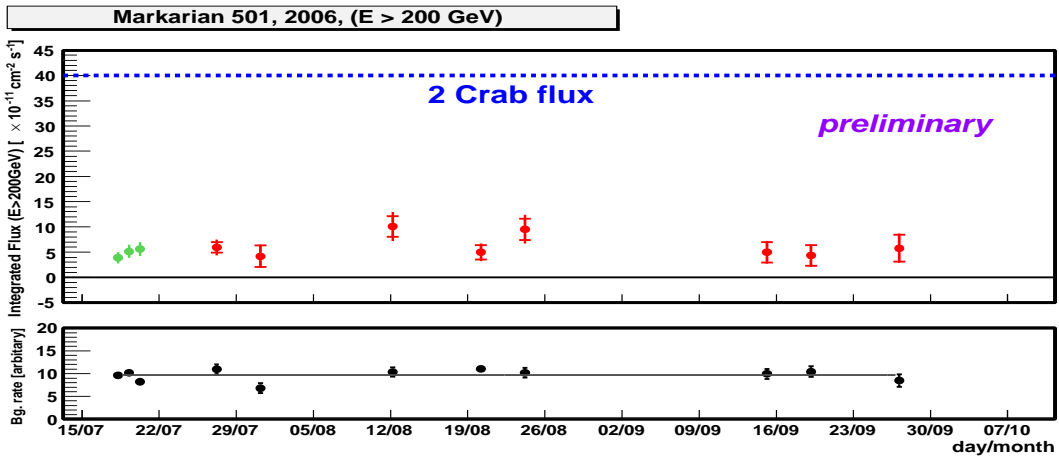}
  \caption{Light curves of Mrk501 observed between July
    and October 2006}\label{fig:Mrk501_lc}
\end{figure*}

\begin{figure*}
  \centering
  \includegraphics[width=14cm,height=6cm,angle=0]{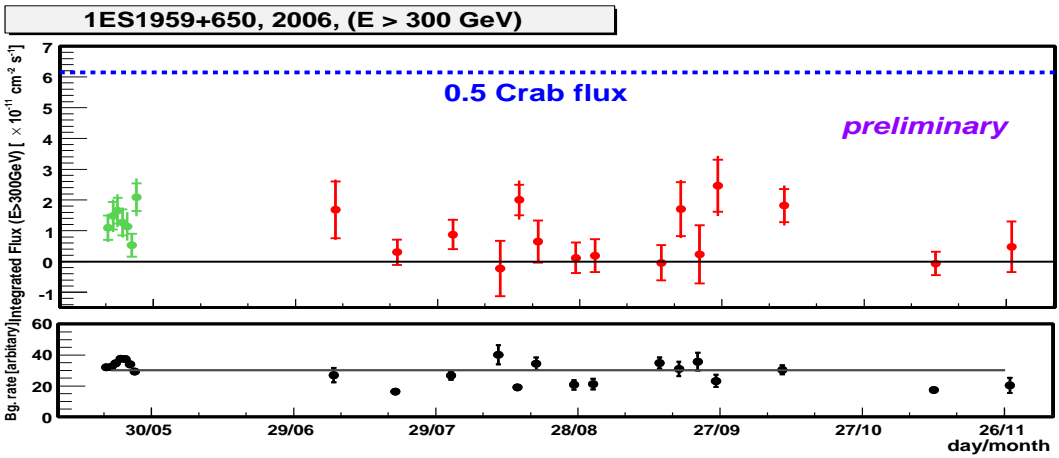}
  \caption{Preliminary light curves of 1ES1959 observed between May
    and November 2006}\label{fig:1ES1959_lc}
\end{figure*}

Preliminary results of the blazar monitoring data recorded
between April 2006 and January 2007 are displayed in
Figures~\ref{fig:Mrk421_lc}, \ref{fig:Mrk501_lc} and
\ref{fig:1ES1959_lc}. The plots show the light curves of the Mrk421,
Mrk501 and 1ES1959 during the period observed with the monitoring
program. The data have been processed with the standard MAGIC analysis
tools. Some observation days have been removed due to poor
observation conditions. On the other hand data taken during
multiwavelength observations~\cite{MultiWavelength} have been
included in the plots as green points. These observations had been scheduled in
advance and can therefore also be considered as unbiased
measurements. Due to longer observation times the errors on the flux
level are however much smaller. The background rate for each night are
shown in the lower plots. 
A flaring state may be defined as 2 times the Crab flux in the case of
Mrk421 and Mrk501 and 0.5 times the Crab flux in case of 1ES1959 as
indicated in the plots. According to this definition Mrk 421 was
observed in flaring state during the commissioning phase of monitoring
program in April/May 2006. 
A statistical analysis of the flux level based on the above data is
ongoing.

\section{Acknowledgments}

We would like to thank the IAC for excellent working conditions. The
support of the German BMBF and MPG, the Italian INFN and the Spanish
CICYT, the Swiss ETH and the Polish MNiI is gratefully acknowledged.

\bibliography{library}
\bibliographystyle{plain}
\end{document}